\documentclass[10pt]{article}
\usepackage{latexsym}
\newcommand{\be}{\begin{equation}}
\newcommand{\ee}{\end{equation}}
\def\n{\noindent}
\catcode `\@=11
\catcode `\@=12
\begin{document}
\begin{center}
\large{\bf {Inhomogeneous Perfect Fluid Universe with Electromagnetic Field
in Lyra Geometry}}\\
\vspace{10mm}
\normalsize{Anirudh Pradhan\footnote{Corresponding author} and Priya Mathur$^2$} \\
\vspace{5mm}
\normalsize{$^1$ Department of Mathematics, Hindu
P. G. College, Zamania-232 331, Ghazipur, U. P., India.} \\
\normalsize{E-mail : acpradhan@yahoo.com, pradhan@iucaa.ernet.in}\\
\vspace{5mm} \normalsize{$^2$ Department of Mathematics,
Institute of Engineering and Technology, Alwar-301 001, Rajasthan, India.} \\
\normalsize{E-mail : priya\_mathur79@yahoo.com}\\
\end{center}
\vspace{10mm}
%\date{}
%\maketitle
\begin{abstract}
A new class of cylindrically symmetric inhomogeneous cosmological
models for perfect fluid distribution with electromagnetic field is
obtained in the context of Lyra's geometry. We have obtained two
types of solutions by considering the uniform as well as time
dependent displacement field. The source of the magnetic field is
due to an electric current produced along the z-axis.Only $F_{12}$
is a non-vanishing component of electromagnetic field tensor. To get
the deterministic solution, it has been assumed that the expansion
$\theta$ in the model is proportional to the shear $\sigma$. It has
been found that the solutions are consistent with the recent
observations of type Ia supernovae and the displacement vector
$\beta(t)$ affects entropy. Physical and geometric aspects of the
models are also discussed in presence and absence of magnetic field.
\end{abstract}
\smallskip
\n PACS: {98.80.Jk, 98.80.-k}  \\
\n Keywords: {Cosmology; cylindrically symmetric universe;
inhomogeneous models; Lyra's geometry }
%\newpage

%%%%%%%%%%%%%%%%%%%%%%%%%%%%%%%%%%%%%%%%%%%%%%%%%%%%%%%%%%%%%%%%%%%%%%%%%%
%%%%%%%%%%%%%%%%%%%%%%%%%%%%%%%   SECTION 1  %%%%%%%%%%%%%%%%%%%%%%%%%%%%%
\section{Introduction and Motivations}
The inhomogeneous cosmological models play a significant role in
understanding some essential features of the universe such as the
formation of galaxies during the early stages of evolution and
process of homogenization. The early attempts at the construction of
such models have been done by Tolman \cite{ref1} and Bondi
\cite{ref2} who considered spherically symmetric models.
Inhomogeneous plane-symmetric models were considered by Taub
\cite{ref3,ref4} and later by Tomimura \cite{ref5}, Szekeres
\cite{ref6}, Collins and Szafron \cite{ref7}, Szafron and Collins
\cite{ref8}. Senovilla \cite{ref9} obtained a new class of exact
solutions of Einstein's equations without big bang singularity,
representing a cylindrically symmetric, inhomogeneous cosmological
model filled with perfect fluid which is smooth and regular
everywhere satisfying energy and causality conditions. Later, Ruiz
and Senovilla \cite{ref10} have examined a fairly large class of
singularity free models through a comprehensive study of general
cylindrically symmetric metric with separable function of $r$ and
$t$ as metric coefficients. Dadhich et al. \cite{ref11} have
established a link between the FRW model and the singularity free
family by deducing the latter through a natural and simple
in-homogenization and anisotropization of the former. Recently,
Patel et al. \cite{ref12} have presented a general class of
inhomogeneous cosmological models filled with non-thermalized
perfect fluid assuming that the background space-time admits two
space-like commuting Killing vectors and has separable metric
coefficients. Singh, Mehta and Gupta \cite{ref13} obtained
inhomogeneous cosmological models of perfect fluid distribution with
electro-magnetic field. Recently, Pradhan et al. \cite{ref14} have
investigated cylindrically-symmetric inhomogeneous cosmological
models in various contexts.
\par
The occurrence of magnetic field on galactic scale is a
well-established fact today, and its importance for a variety of
astrophysical phenomena is generally acknowledged as pointed out by
Zeldovich et al. \cite{ref15}. Also Harrison \cite{ref16} suggests
that magnetic field could have a cosmological origin. As a natural
consequences, we should include magnetic fields in the
energy-momentum tensor of the early universe. The choice of
anisotropic cosmological models in Einstein system of field
equations leads to the cosmological models more general than
Robertson-Walker model \cite{ref17}. The presence of primordial
magnetic field in the early stages of the evolution of the universe
is discussed by many \cite{ref18}$-$\cite{ref27}. Strong magnetic
field can be created due to adiabatic compression in clusters of
galaxies. Large-scale magnetic field gives rise to anisotropies in
the universe. The  anisotropic pressure created by the magnetic
fields dominates the evolution of the shear anisotropy and decays
slowly as compared to the case when the pressure is held isotropic
\cite{ref28,ref29}. Such fields can be generated at the end of an
inflationary epoch \cite{ref30}$-$\cite{ref34}. Anisotropic magnetic
field models have significant contribution in the evolution of
galaxies and stellar objects. Bali and Ali \cite{ref35} obtained a
magnetized cylindrically symmetric universe with an electrically
neutral perfect fluid as the source of matter. Pradhan et al.
\cite{ref36} have investigated magnetized cosmological models in
various contexts.
\par
In 1917 Einstein introduced the cosmological constant into his field
equations of general relativity in order to obtain a static
cosmological model since, as is well known, without the cosmological
term his field equations admit only non-static solutions. After the
discovery of the red-shift of galaxies and explanation thereof
Einstein regretted for the introduction of the cosmological
constant. Recently, there has been much interest in the cosmological
term in context of quantum field theories, quantum gravity,
super-gravity theories, Kaluza-Klein theories and the
inflationary-universe scenario. Shortly after Einstein's general
theory of relativity Weyl \cite{ref37} suggested the first so-called
unified  field theory based on a generalization of Riemannian
geometry. With its backdrop, it would seem more appropriate to call
Weyl's theory a geometrized theory of gravitation and
electromagnetism (just as the general theory was a geometrized
theory of gravitation only), instead a unified field theory. It is
not clear as to what extent the two fields have been unified, even
though they acquire (different) geometrical significance in the same
geometry. The theory was never taken seriously inasmuchas it was
based on the concept of non-integrability of length transfer; and,
as pointed out by Einstein, this implies that spectral frequencies
of atoms depend on their past histories and therefore have no
absolute significance. Nevertheless, Weyl's geometry provides an
interesting example of non-Riemannian connections, and recently
Folland \cite{ref38} has given a global formulation of Weyl
manifolds clarifying considerably many of Weyl's basic ideas
thereby.
\newline
\par
In 1951 Lyra \cite{ref39} proposed a modification of Riemannian
geometry by introducing a gauge function into the structure-less
manifold, as a result of which the cosmological constant arises
naturally from the geometry. This bears a remarkable resemblance to
Weyl's geometry. But in Lyra's geometry, unlike that of Weyl, the
connection is metric preserving as in Remannian; in other words,
length transfers are integrable. Lyra also introduced the notion of
a gauge and in the ``normal'' gauge the curvature scalar in
identical to that of Weyl. In consecutive investigations Sen
\cite{ref40}, Sen and Dunn \cite{ref41}  proposed a new
scalar-tensor theory of gravitation and constructed an analog of the
Einstein field equations based on Lyra's geometry. It is, thus,
possible \cite{ref40} to construct a geometrized theory of
gravitation and electromagnetism much along the lines of Weyl's
``unified'' field theory, however, without the inconvenience of
non-integrability length transfer.
\newline
\par
Halford \cite{ref42} has pointed out that the constant vector
displacement field $\phi_i$ in Lyra's geometry plays the role of
cosmological constant $\Lambda$ in the normal general relativistic
treatment. It is shown by Halford \cite{ref43} that the
scalar-tensor treatment based on Lyra's geometry predicts the same
effects within observational limits as the Einstein's theory.
Several authors Sen and Vanstone \cite{ref44}, Bhamra \cite{ref45},
Karade and Borikar \cite{ref46}, Kalyanshetti and Wagmode
\cite{ref47}, Reddy and Innaiah \cite{ref48}, Beesham \cite{ref49},
Reddy and Venkateswarlu \cite{ref50}, Soleng \cite{ref51}, have
studied cosmological models based on Lyra's manifold with a constant
displacement field vector. However, this restriction of the
displacement field to be constant is merely one for convenience and
there is no {\it a priori} reason for it. Beesham \cite{ref52}
considered FRW models with time dependent displacement field. He has
shown that by assuming the energy density of the universe to be
equal to its critical value, the models have the $k=-1$ geometry.
Singh and Singh \cite{ref53}$-$ \cite{ref56}, Singh and Desikan
\cite{ref57} have studied Bianchi-type I, III, Kantowaski-Sachs and
a new class of cosmological models with time dependent displacement
field and have made a comparative study of Robertson-Walker models
with constant deceleration parameter in Einstein's theory with
cosmological term and in the cosmological theory based on Lyra's
geometry. Soleng \cite{ref51} has pointed out that the cosmologies
based on Lyra's manifold with constant gauge vector $\phi$ will
either include a creation  field and be equal to Hoyle's creation
field cosmology \cite{ref56}$-$ \cite{ref60} or contain a special
vacuum field, which together with the gauge vector term, may be
considered as a cosmological term. In the latter case the solutions
are equal to the general relativistic cosmologies with a
cosmological term.
\newline
\par
Recently, Pradhan et al. \cite{ref61}, Casama et al. \cite{ref62},
Rahaman et al. \cite{ref63}, Bali and Chandani \cite{ref64}, Kumar
and Singh \cite{ref65}, Singh \cite{ref66} and Rao, Vinutha and
Santhi \cite{ref67} have studied cosmological models based on Lyra's
geometry in various contexts. With these motivations, in this paper,
we have obtained exact solutions of Einstein's field equations in
cylindrically symmetric inhomogeneous space-time within the frame
work of Lyra's geometry in the presence and absence of magnetic
field for uniform and time varying displacement vector. This paper
is organized as follows. In Section $1$ the motivation for the
present work is discussed. The metric and the field equations are
presented in Section $2$, in Section $3$ the solution of field
equations, the Section $4$ contains the solution of uniform
displacement field ($\beta = \beta_{0}$, constant). The Section 5
deals with the solution with time varying displacement field ($\beta
= \beta(t)$). Subsections $5.1, 5.2$ and $5.3$ describe the
solutions of Empty Universe, Zeldovich Universe and Radiating
Universe with the physical and geometric aspects of the models
respectively. The solutions in absence of magnetic field are given
in Section $6$. Sections $7$ and $8$ deal with the solutions for
uniform and time dependent displacement field. Finally, in Section
$9$ discussion and concluding remarks are given.
%%%%%%%%%%%%%%%%%%%%%%%%%%%%%%%%%%%%%%%%%%%%%%%%%%%%%%%%%%%%%%%%%%%%%%%%%%
%%%%%%%%%%%%%%%%%%%%%%%%%%%%%%%%%%%%%% SECTION 2 %%%%%%%%%%%%%%%%%%%%%%%%%%
\section{The Metric and Field  Equations}
We consider the cylindrically symmetric metric in the form
\begin{equation}
\label{eq1}
ds^{2} = A^{2}(dx^{2} - dt^{2}) + B^{2} dy^{2} + C^{2} dz^{2},
\end{equation}
where $A$ is the function of $t$ alone and  $B$ and $C$ are functions of $x$ and $t$.
The energy momentum tensor is taken as has the form
\begin{equation}
\label{eq2}
T^{j}_{i} = (\rho + p)u_{i} u^{j} + p g^{j}_{i} +  E^{j}_{i},
\end{equation}
where $\rho$ and $p$ are, respectively, the energy density and
pressure of the cosmic fluid,and $u_{i}$ is the fluid four-velocity
vector satisfying the condition
\begin{equation}
\label{eq3}
u^{i} u_{i} = -1, ~ ~  u^{i} x_{i} = 0.
\end{equation}
In Eq. (\ref{eq2}), $E^{j}_{i}$ is the electromagnetic field given
by Lichnerowicz \cite{ref68}
\begin{equation}
\label{eq4}
E^{j}_{i} = \bar{\mu}\left[h_{l}h^{l}\left(u_{i}u^{j} + \frac{1}{2}g^{j}_{i}\right)
- h_{i}h^{j}\right],
\end{equation}
where $\bar{\mu}$ is the magnetic permeability and $h_{i}$ the magnetic flux vector
defined by
\begin{equation}
\label{eq5}
h_{i} = \frac{1}{\bar{\mu}} \, {^*}F_{ji} u^{j},
\end{equation}
where the dual electromagnetic field tensor $^{*}F_{ij}$ is defined
by Synge \cite{ref69}
\begin{equation}
\label{eq6}
^{*}F_{ij} = \frac{\sqrt{-g}}{2} \epsilon_{ijkl} F^{kl}.
\end{equation}
Here $F_{ij}$ is the electromagnetic field tensor and $\epsilon_{ijkl}$ is the Levi-Civita
tensor density.\\
The co-ordinates are considered to be co-moving so that $u^{1}$ =
$0$ = $u^{2}$ = $u^{3}$ and $u^{4} = \frac{1}{A}$. If we consider
that the current flows  along the $z$-axis, then $F_{12}$ is the
only non-vanishing component of $F_{ij}$. The Maxwell's equations
\begin{equation}
\label{eq7}
F_[ij;k] = 0,
\end{equation}
\begin{equation}
\label{eq8}
\left[\frac{1}{\bar{\mu}}F^{ij}\right]_{;j} = 4 \pi J^{i},
\end{equation}
require that $F_{12}$ is the function of x-alone. We assume that the
magnetic permeability is the functions of $x$ and $t$ both. Here the
semicolon represents a covariant
differentiation\\

The field equations (in gravitational units $c = 1$, $G = 1$), in normal 
gauge for Lyra's manifold, obtained by
Sen \cite{ref4} as
\begin{equation}
\label{eq9}
R_{ij} - \frac{1}{2} g_{ij} R + \frac{3}{2} \phi_i \phi_j
- \frac{3}{4} g_{ij} \phi_k \phi^k = - 8 \pi T_{ij},
\end{equation}
where $\phi_{i}$ is the displacement field vector defined as
\begin{equation}
\label{eq10} \phi_{i} = (0, 0, 0, \beta),
\end{equation}
where $\beta$ is either a constant or a function of $t$. The other symbols have
their usual meaning as in Riemannian geometry. \\

For the line-element (\ref{eq1}), the field Eq. (\ref{eq9}) with
Eqs. (\ref{eq2}) and (\ref{eq10}) lead to the following system of
equations
\[
\frac{1}{A^{2}}\left[- \frac{B_{44}}{B} - \frac{C_{44}}{C} +
\frac{A_{4}}{A} \left(\frac{B_{4}}{B} +\frac{C_{4}}{C}\right) -
\frac{B_{4} C_{4}}{B C} + \frac{B_{1}C_{1}}{BC}\right] -
\frac{3}{4}\beta^{2}
\]
 \begin{equation}
\label{eq11}
= 8 \pi \left(p + \frac{F^{2}_{12}}{2\bar{\mu} A^{2} B^{2}} \right),
\end{equation}
\begin{equation}
\label{eq12}
\frac{1}{A^{2}}\left(\frac{A^{2}_{4}}{A^{2}}- \frac{A_{44}}{A} - \frac{C_{44}}{C} +
\frac{C_{11}}{C} \right) - \frac{3}{4}\beta^{2}   =  8 \pi \left(p + \frac{F^{2}_{12}}
{2\bar{\mu} A^{2} B^{2}} \right),
\end{equation}
\begin{equation}
\label{eq13}
\frac{1}{A^{2}}\left(\frac{A^{2}_{4}}{A^{2}} - \frac{A_{44}}{A} - \frac{B_{44}}{B} +
\frac{B_{11}}{ B}\right) - \frac{3}{4}\beta^{2}   =  8 \pi \left(p - \frac{F^{2}_{12}}
{2\bar{\mu} A^{2} B^{2}} \right),
\end{equation}
\[
\frac{1}{A^{2}}\left[- \frac{B_{11}}{B} - \frac{C_{11}}{C} +
\frac{A_{4}}{A} \left(\frac{B_{4}}{B} +\frac{C_{4}}{C}\right) -
\frac{B_{1}C_{1}}{BC}  + \frac{B_{4} C_{4}}{B C}\right] +
\frac{3}{4}\beta^{2}
\]
\begin{equation}
\label{eq14}
= 8 \pi \left(\rho + \frac{F^{2}_{12}}{2\bar{\mu} A^{2} B^{2}}\right),
\end{equation}
\begin{equation}
\label{eq15}
\frac{B_{14}}{B} + \frac{C_{14}}{C} - \frac{A_{4}}{A}\left(\frac{B_{1}}{B} +
\frac{C_{1}}{C}\right) = 0,
\end{equation}
where the subscript indices $1$ and $4$ in A, B, C and elsewhere
denote ordinary differentiation with respect to $x$ and $t$
respectively.
%%%%%%%%%%%%%%%%%%%%%%%%%%%%%%%%%%%%%%%%%%%%%%%%%%%%%%%%%%%%%%%%%%%%%%%%%%%%%%%%%%%
%%%%%%%%%%%%%%%%%%%%%%%%%%%%%%%  SECTION 3  %%%%%%%%%%%%%%%%%%%%%%%%%%%%%%%%%%%%%%%%
\section{Solution of Field Equations}
Equations (\ref{eq11})-(\ref{eq15}) are five independent equations
in seven unknowns $A$, $B$, $C$, $\rho$, $p$, $\beta$ and $F_{12}$.
For the complete determinacy of the system, we need two extra
conditions which are narrated hereinafter. The research on exact
solutions is based on some physically reasonable restrictions used
to simplify
the field equations. \\

To get determinate solution we assume that the expansion $\theta$ in
the model is proportional to the shear $\sigma$. This condition
leads to
\begin{equation}
\label{eq16}
A = \left(\frac{B}{C}\right)^{n},
\end{equation}
where $n$ is a constant. The motive behind assuming this condition
is explained with reference to Thorne \cite{ref70}, the observations
of the velocity-red-shift relation for extragalactic sources suggest
that Hubble expansion of the universe is isotropic today within
$\approx 30$ per cent \cite{ref71,ref72}. To put more precisely,
red-shift studies place the limit
$$
\frac{\sigma}{H} \leq 0.3
$$
on the ratio of shear, $\sigma$, to Hubble constant, $H$, in the
neighbourhood of our Galaxy today. Collins et al. \cite{ref73} have
pointed out that for spatially homogeneous metric, the normal
congruence to the
homogeneous expansion satisfies that the condition $\frac{\sigma}{\theta}$ is constant. \\

From Eqs. (\ref{eq11})-(\ref{eq13}), we have
\begin{equation}
\label{eq17}
\frac{A_{44}}{A} - \frac{A^{2}_{4}}{A^{2}} +
\frac{A_{4}B_{4}}{AB} + \frac{A_{4}C_{4}} {AC} -\frac{B_{44}}{B} -
\frac{B_{4}C_{4}}{BC}  = \frac{C_{11}}{C} - \frac{B_{1}C_{1}}{BC} =
\mbox{K (constant)}
\end{equation}
and
\begin{equation}
\label{eq18}
\frac{8\pi F^{2}_{12}}{\bar{\mu}B^{2}} = -
\frac{C_{44}}{C} + \frac{C_{11}}{C} + \frac{B_{44}}{B} -
\frac{B_{11}}{B}.
\end{equation}
We also assume that
\[
B = f(x)g(t),
\]
\begin{equation}
\label{eq19}
C = f(x)k(t).
\end{equation}
Using Eqs. (\ref{eq16}) and (\ref{eq19}) in (\ref{eq15}) and (\ref{eq17}) lead to
\begin{equation}
\label{eq20}
\frac{k_{4}}{k} = \frac{(2n - 1)}{(2n + 1)}\frac{g_{4}}{g},
\end{equation}
\begin{equation}
\label{eq21}
(n - 1)\frac{g_{44}}{g} - n\frac{k_{44}}{k} - \frac{g_{4}}{g}\frac{k_{4}}{k} = K,
\end{equation}
\begin{equation}
\label{eq22}
f f_{11} - f^{2}_{1} = Kf^{2}.
\end{equation}
Equation (\ref{eq20}) leads to
\begin{equation}
\label{eq23}
k = cg^{\alpha},
\end{equation}
where $\alpha = \frac{2n - 1}{2n + 1}$ and $c$ is the constant of
integration. From Eqs. (\ref{eq21}) and (\ref{eq23}), we have
\begin{equation}
\label{eq24}
\frac{g_{44}}{g} + \ell \frac{g^{2}_{4}}{g^{2}} = N,
\end{equation}
where
$$
\ell = \frac{n\alpha(\alpha - 1) + \alpha}{n(\alpha - 1) + 1}, \, \, N =
\frac{K}{n(1 - \alpha) - 1}.
$$
Equation (\ref{eq22}) leads to
\begin{equation}
\label{eq25}
f = \exp{\left(\frac{1}{2}K(x + x_{0})^{2}\right)},
\end{equation}
where $x_{0}$ is an integrating constant. Equation (\ref{eq24}) leads to
\begin{equation}
\label{eq26}
g = \left(c_{1}e^{bt} + c_{2}e^{-bt}\right)^{\frac{1}{(\ell + 1)}},
\end{equation}
where $b = \sqrt{(\ell + 1)N}$ and $c_{1}$, $c_{2}$ are integrating
constants. Hence from (\ref{eq23}) and (\ref{eq26}), we have
\begin{equation}
\label{eq27}
k = c\left(c_{1}e^{bt} + c_{2}e^{-bt}\right)^{\frac{\alpha}{(\ell + 1)}}.
\end{equation}
Therefore we obtain
\begin{equation}
\label{eq28}
B = \exp{\left(\frac{1}{2}K(x + x_{0})^{2}\right)} \left(c_{1}e^{bt} + c_{2}e^{-bt}
\right)^{\frac{1}{(\ell + 1)}},
\end{equation}
\begin{equation}
\label{eq29}
C = \exp{\left(\frac{1}{2}K(x + x_{0})^{2}\right)} c \left(c_{1}e^{bt} + c_{2}e^{-bt}
\right)^{\frac{\alpha}{(\ell + 1)}},
\end{equation}
\begin{equation}
\label{eq30}
A = a \left(c_{1}e^{bt} + c_{2}e^{-bt}\right)^{\frac{n(1 - \alpha)}{(\ell + 1)}},
\end{equation}
where $a = \frac{c_{3}}{c}$, $c_{3}$ being a constant of integration. \\

After using suitable transformation of the co-ordinates, the model (\ref{eq1}) reduces
to the form
\[
ds^{2} = a^{2}(c_{1}e^{bT} + c_{2}e^{-bT})^{\frac{2n(1 -
\alpha)}{(\ell + 1)}} (dX^{2} - dT^{2}) + e^{KX^{2}}(c_{1}e^{bT} +
c_{2}e^{-bT})^{\frac{2}{(\ell + 1)}} dY^{2}
\]
\begin{equation}
\label{eq31}
+ e^{KX^{2}}(c_{1}e^{bT} + c_{2}e^{-bT})^{\frac{2\alpha}{(\ell + 1)}} dZ^{2},
\end{equation}
where $x + x_{0} = X$, $t = T$, $y = Y$, $cz = Z$. \\

For the specification of displacement vector $\beta$ within the
framework of Lyra geometry and for realistic models of physical
importance, we consider following two cases described in Sections $4$ and $5$.
%%%%%%%%%%%%%%%%%%%%%%%%%%%%%%%%%%%%%%%%%%%%%%%%%%%%%%%%%%%%%%%%%%%%%%%%%%%%%%%%%%%
%%%%%%%%%%%%%%%%%%%%%%%%%%%%%%%  SECTION 4  %%%%%%%%%%%%%%%%%%%%%%%%%%%%%%%%%%%%%%%%
\section{When $\beta$ is a constant i.e. $\beta = \beta_{0}$ (constant)}
Using Eqs. (\ref{eq28}), (\ref{eq29}) and (\ref{eq30}) in Eqs. (\ref{eq11}) and
(\ref{eq14}) the expressions for
pressure $p$ and density $\rho$ for the model (\ref{eq31}) are given by
\[
8 \pi p = \frac{1}{a^{2}\psi_{2}^{\frac{2n(1 - \alpha)}{(\ell + 1)}}}
\Biggl[K^{2}X^{2} - \frac{2(3 + \alpha)b^{2}c_{1}c_{2}}
{(\ell + 1)\psi_{2}^{2}}
\]
\begin{equation}
\label{eq32}
- \frac{(2n \alpha^{2} + \alpha^{2} + 2\alpha - 2n + 3)b^{2}}{2(\ell + 1)^{2}}
\frac{\psi_{1}^{2}}{\psi_{2}^{2}}\Biggr]
- \frac{3}{4}\beta_{0}^{2},
\end{equation}
\[
8 \pi \rho = \frac{1}{a^{2}\psi_{2}^{\frac{2n(1 - \alpha)}{(\ell +
1)}}} \Biggl[- 3 K^{2}X^{2} - 2K + \frac{2b^{2} (\alpha - 1)c_{1}c_{2}}{(\ell +
1)\psi_{2}^{2}} 
\]
\begin{equation}
\label{eq33}
- \frac{(2n \alpha^{2} - \alpha^{2} - 2\alpha - 2n + 1)b^{2}}{2(\ell + 1)^{2}}
\frac{\psi_{1}^{2}}{\psi_{2}^{2}}\Biggr]
+ \frac{3}{4}\beta_{0}^{2},
\end{equation}
where
\[
\psi_{1} = c_{1}e^{bT} - c_{2}e^{-bT},
\]
\[
\psi_{2} = c_{1}e^{bT} + c_{2}e^{-bT}.
\]
From Eq. \ref{eq18}) the non-vanishing component $F_{12}$ of the
electromagnetic field tensor is obtained as
\begin{equation}
\label{eq34}
F_{12}^{2} = \frac{\bar{\mu}}{8\pi}\frac{b^{2}(1 - \alpha)}{(\ell + 1)^{2}}e^{KX^{2}}
\psi_{2}^{\frac{2}{(\ell + 1)}}
\Biggl[\frac{4(\ell + 1)c_{1}c_{2} + (1 + \alpha)\psi_{1}^{2}}{\psi_{2}^{2}}\Biggr].
\end{equation}
From above equation it is observed that the electromagnetic field tensor increases
with time.\\

The reality conditions (Ellis \cite{ref74})
$$
(i) \rho + p > 0, ~ ~ (ii) \rho + 3p > 0,
$$
lead to
\begin{equation}
\label{eq35}
\frac{b^{2}(n - n\alpha^{2} - 1)}{(\ell + 1)^{2}}\frac{\psi_{1}^{2}}{\psi_{2}^{2}} -
\frac{4b^{2}c_{1}c_{2}}{(\ell + 1)\psi_{2}^{2}} > K (KX^{2} + 1),
\end{equation}
and
\[
\frac{b^{2}(4n - 4n\alpha^{2} - \alpha^{2} - 2\alpha - 5)}{(\ell + 1)^{2}}
\frac{\psi_{1}^{2}}{\psi_{2}^{2}} - \frac{4b^{2}(\alpha + 5)c_{1}c_{2}}
{(\ell + 1)\psi_{2}^{2}} 
\]
\begin{equation}
\label{eq36}
> 2K + \frac{3}{2}\beta_{0}^{2}a^{2}
\psi_{2}^{\frac{2n(1 - \alpha)}{(\ell + 1)}},
\end{equation}
respectively.

The dominant energy conditions (Hawking and Ellis \cite{ref75})
$$
(i) \rho - p \geq 0, ~ ~ (ii) \rho + p \geq 0,
$$
lead to
\begin{equation}
\label{eq37}
\frac{b^{2}(\alpha + 1)^{2}}{(\ell + 1)^{2}}\frac{\psi_{1}^{2}}{\psi_{2}^{2}} +
\frac{4b^{2}(\alpha + 1)c_{1}c_{2}}{(\ell + 1)\psi_{2}^{2}}  + \frac{3}{2}
\beta_{0}^{2}a^{2}\psi_{2}^{\frac{2n(1 - \alpha)}
{(\ell + 1)}} \geq  2K(2K X^{2} + 1),
\end{equation}
and
\begin{equation}
\label{eq38}
\frac{b^{2}(n - n\alpha^{2} - 1)}{(\ell + 1)^{2}}\frac{\psi_{1}^{2}}{\psi_{2}^{2}}
- \frac{4b^{2}c_{1}c_{2}}{(\ell + 1)\psi_{2}^{2}} \geq K(KX^{2} + 1),
\end{equation}
respectively. The conditions (\ref{eq36}) and (\ref{eq37}) impose a restriction 
on constant displacement vector $\beta_{0}$.
%%%%%%%%%%%%%%%%%%%%%%%%%%%%%%%%%%%%%%%%%%%%%%%%%%%%%%%%%%%%%%%%%%%%%%%%%%%%%%%%%%%
%%%%%%%%%%%%%%%%%%%%%%%%%%%%%%%  SECTION 5  %%%%%%%%%%%%%%%%%%%%%%%%%%%%%%%%%%%%%%%%
\section{When $\beta$ is a function of $t$ i.e. $\beta$ = $\beta(t)$}
In this case to find the explicit value of displacement field $\beta(t)$, we assume
that the fluid obeys an
equation of state of the form
\begin{equation}
\label{eq39}
p = \gamma \rho,
\end{equation}
where $\gamma(0 \leq \gamma \leq 1)$ is a constant. Using Eqs. (\ref{eq28}) - (\ref{eq30}) 
and (\ref{eq39}) in equations (\ref{eq11}) and (\ref{eq14}) we obtain
\begin{equation}
\label{eq40}
4 \pi(1 + \gamma)\rho = \frac{1}{a^{2}\psi_{2}^{\frac{2n(1 - \alpha)}{(\ell +
1)}}} \Biggl[- K^{2}X^{2} - K - \frac{4b^{2}c_{1}c_{2}}{(\ell + 1)\psi_{2}^{2}} 
- \frac{b^{2}(n - n\alpha^{2} - 1)}{(\ell + 1)^{2}}\frac{\psi_{1}^{2}}{\psi_{2}^{2}}\Biggr],
\end{equation}  
and
\[
(1 + \gamma)\beta^{2}{(t)} = \frac{4}{3a^{2}\psi_{2}^{\frac{2n(1 - \alpha)}
{(\ell + 1)}}}\Biggl[K^{2}X^{2}(1 + \gamma) + 2 K\gamma  
\]
\[
+ \, \frac{2b^{2}c_{1}c_{2}\{(1 - \alpha)(1 - \gamma) - 4 \}}{(\ell + 1)\psi_{2}^{2}} 
\]
\begin{equation}
\label{eq41}
+ \, \frac{b^{2}}{(\ell + 1)^{2}}\{(2n\alpha^{2} - \alpha^{2} - 2\alpha - 2n + 1)(1 + \gamma)
- 2(n \alpha^{2} - n + 1)\}\frac{\psi_{1}^{2}}{\psi_{2}^{2}}\Biggr].
\end{equation}
Here we consider the three cases of physical interest in following 
Subsections $5.1$, $5.2$ and $5.3$. 
%%%%%%%%%%%%%%%%%%%%%%%%%%%%%%%  SUBSECTION 5.1  %%%%%%%%%%%%%%%%%%%%%%%%%%%%%%%%%%%%%%%%
\subsection{Empty Universe}
Putting $\gamma = 0$ in (\ref{eq39}) reduces to $p = 0$. Thus, from Eqs.
(\ref{eq40}) and (\ref{eq41}), we obtain the expressions for physical 
parameters $\rho$ and $\beta^{2}{(t)}$ 
\begin{equation}
\label{eq42}
4 \pi \rho = \frac{1}{a^{2}\psi_{2}^{\frac{2n(1 - \alpha)}{(\ell +
1)}}} \Biggl[- K^{2}X^{2} - K - \frac{4b^{2}c_{1}c_{2}}{(\ell + 1)
\psi_{2}^{2}} + \frac{b^{2}(n - n\alpha^{2} - 1)}{(\ell + 1)^{2}}
\frac{\psi_{1}^{2}}{\psi_{2}^{2}}\Biggr],
\end{equation} 
\begin{equation}
\label{eq43}
\beta^{2}(t) = \frac{4}{3a^{2}\psi_{2}^{\frac{2n(1 - \alpha)}
{(\ell + 1)}}}\Biggl[K^{2} X^{2} - \frac{2b^{2}(\alpha + 4)c_{1}c_{2}}
{(\ell + 1)\psi_{2}^{2}} - \frac{b^{2} (\alpha + 1)^{2}}{(\ell + 1)^{2}}
\frac{\psi_{1}^{2}}{\psi_{2}^{2}}\Biggr].
\end{equation}
From Eqs. (\ref{eq42}) and (\ref{eq43}), we observe that $\rho > 0$ and 
$\beta^{2}{(t)} > 0$ according as
\begin{equation}
\label{eq44}
\frac{b^{2}(n - n\alpha^{2} - 1)}{(\ell + 1)^{2}}\psi_{1}^{2} - K(KX^{2} + 1)
\psi_{2}^{2} > \frac{4b^{2}c_{1}c_{2}}{(\ell + 1)},
\end{equation} 
and
\begin{equation}
\label{eq45}
K^{2}X^{2}\psi_{2}^{2} - \frac{b^{2}(\alpha + 1)^{2}}{(\ell + 1)^{2}}\psi_{1}^{2} 
> \frac{2b^{2}(\alpha + 4)c_{1}c_{2}}{(\ell + 1)},
\end{equation}
respectively.

Halford \cite{ref6} has pointed out that the displacement field 
$\phi_i$ in Lyra's geometry plays the role of cosmological constant 
$\Lambda$ in the normal general relativistic treatment. From Eq. 
(\ref{eq43}), it is observed that the displacement vector $\beta(t)$ 
is a decreasing function of time which is corroborated with Halford 
as well as with the recent observations \cite{ref76,ref77} leading to 
the conclusion that $\Lambda(t)$ is a decreasing function of $t$.
%%%%%%%%%%%%%%%%%%%%%%%%%%%%%%%  SUBSECTION 5.2  %%%%%%%%%%%%%%%%%%%%%%%%%%%%%%%%%
%%%%%%%%%%%%%%%%%%%%%%%%%%%%%%%%%%%%%%%%%%%%%%%%%%%%%%%%%%%%%%%%%%%%%%%%%%%%%%%%%
\subsection{Zeldovich Universe}
Putting $\gamma = 1$ in Eq. (\ref{eq39}) reduces to $\rho = p$. In this case 
the expressions for physical quantities are given by
\begin{equation}
\label{eq46}
\beta^{2}(t) = \frac{4}{3a^{2}\psi_{2}^{\frac{2n(1 - \alpha)}{(\ell + 1)}}}
\Biggl[K^{2} X + K - \frac{4b^{2}c_{1}c_{2}}{(\ell + 1)\psi_{2}^{2}} -
\frac{b^{2}(\alpha + 1)^{2}}{(\ell + 1)^{2}}\frac{\psi_{1}^{2}}{\psi_{2}^{2}}\Biggr].
\end{equation}
\[
8\pi p = 8\pi \rho = \frac{1}{a^{2}\psi_{2}^{\frac{2n(1 - \alpha)}{(\ell + 1)}}}
\Biggl[- K^{2} X^{2} - K - \frac{4b^{2}c_{1}c_{2}}
{(\ell + 1)\psi_{2}^{2}} 
\]
\begin{equation}
\label{eq47}
+ \, \frac{b^{2}(n - n\alpha^{2} - 1)}
{(\ell + 1)^{2}}\frac{\psi_{1}^{2}}{\psi_{2}^{2}}\Biggr].
\end{equation}
The reality condition (Ellis \cite{ref74})
$$
(i) \rho + p > 0, ~ ~ (ii) \rho + 3p > 0,
$$
lead to
\begin{equation}
\label{eq48}
\frac{b^{2}(n - n\alpha^{2} - 1)}{(\ell + 1)^{2}}\psi_{1}^{2} - K(KX^{2} + 1)
\psi_{2}^{2} > \frac{4b^{2}c_{1}c_{2}}{(\ell + 1)}.
\end{equation} 
%%%%%%%%%%%%%%%%%%%%%%%%%%%%%%%  SUBSECTION 5.3  %%%%%%%%%%%%%%%%%%%%%%
\subsection{Radiating Universe}
Putting $\gamma = \frac{1}{3}$ in Eq. (\ref{eq39}) reduces to $p = \frac{1}{3}\rho$. 
In this case the expressions for $\beta(t)$, $p$ and $\rho$ are obtained as
\[
\beta^{2}(t) = \frac{2}{3a^{2}\psi_{2}^{\frac{2n(1 - \alpha)}{(\ell + 1)}}}
\Biggl[2K^{2}X^{2} + K + \frac{2 b^{2}(\alpha + 5)c_{1}c_{2}}{(\ell + 1)
\psi_{2}^{2}}
\]
\begin{equation}
\label{eq49}
+ \frac{b^{2}(n\alpha^{2} - 2\alpha^{2} - 4\alpha - n - 1)}{3(\ell + 1)^{2}}
\frac{\psi_{1}^{2}}
{\psi_{2}^{2}}\Biggr],
\end{equation}
\begin{equation}
\label{eq50}
8\pi p = \frac{1}{2a^{2}\psi_{2}^{\frac{2n(1 - \alpha)}{(\ell + 1)}}}
\Biggl[- K^{2} X^{2} - K - \frac{4 b^{2}c_{1}c_{2}}{(\ell + 1)\psi_{2}^{2}} 
+ \frac{b^{2}(n - n\alpha^{2} - 1)}{(\ell + 1)^{2}}\frac{\psi_{1}^{2}}
{\psi_{2}^{2}}\Biggr],
\end{equation}
\begin{equation}
\label{eq51}
8\pi \rho = \frac{3}{2a^{2}\psi_{2}^{\frac{2n(1 - \alpha)}{(\ell + 1)}}}
\Biggl[- K^{2} X^{2} - K - \frac{4 b^{2}c_{1}c_{2}}{(\ell + 1)\psi_{2}^{2}} 
+ \frac{b^{2}(n - n \alpha^{2} - 1)}{(\ell + 1)^{2}}\frac{\psi_{1}^{2}}
{\psi_{2}^{2}}\Biggr].
\end{equation}
From Eq. (\ref{eq49}), it is observed that displacement vector $\beta$ is decreasing
function of time and therefore it behaves as cosmological term $\Lambda$.

The reality conditions (Ellis \cite{ref74})
$$
(i) \rho + p > 0, ~ ~ (ii) \rho + 3p > 0,
$$
are satisfied under condition (\ref{eq48}).

The dominant energy conditions (Hawking and Ellis\cite{ref75})
$$
(i) \rho - p \geq 0, ~ ~ (ii) \rho + p \geq 0,
$$
lead to
\begin{equation}
\label{eq52}
\frac{b^{2}(n - n\alpha^{2} - 1)}{(\ell + 1)^{2}}\psi_{1}^{2} - K(KX^{2} + 1)
\psi_{2}^{2} \geq \frac{4b^{2}c_{1}c_{2}}{(\ell + 1)}.
\end{equation} \\
 
%%%%%%%%%%%%%%%%%%%%%%%%%%%%%%%%%%%%%%%%%%%%%%%%%%%%%%%%%%%%%%%%%%%%%%%%%%%%%
\noindent
{\bf{Some Geometric Properties of the Model}} \\
The expressions for the expansion $\theta$, shear scalar $\sigma^{2}$, deceleration
parameter $q$ and proper
volume $V^{3}$ for the model (\ref{eq31}) are given by
\begin{equation}
\label{eq53}
\theta = \frac{b\{n(1 - \alpha) + (1 + \alpha)\}}{(\ell + 1)a \psi_{2}^
{\frac{n(1 - \alpha)}{(\ell + 1)}}}
\frac{\psi_{1}}{\psi_{2}},
\end{equation}
\begin{equation}
\label{eq54}
\sigma^{2} = \frac{b^{2}\left[\{n(1 - \alpha) + (1 + \alpha)\}^{2}- 3n(1 - \alpha)
(1 + \alpha) - 3\alpha\right]}
{3(\ell + 1)^{2}a^{2} \psi_{2}^{\frac{2n(1 - \alpha)}{(\ell + 1)}}}
\frac{\psi_{1}^{2}}{\psi_{2}^{2}},
\end{equation}
\begin{equation}
\label{eq55}
q = - 1 - \frac{6 c_{1}{c_{2}}(\ell + 1)}{n(1 - \alpha^{2})\left(c_{1}e^{bT}
- c_{2}e^{-bT}\right)^{2}},
\end{equation}
\begin{equation}
\label{eq56}
V^{3} = \sqrt{-g} = a^{2}\psi_{2}^{\frac{2n(1 + \alpha)(1 - \alpha)}
{(\ell + 1)}}e^{KX^{2}}.
\end{equation}
From Eqs. (\ref{eq53}) and (\ref{eq54}) we obtain
\begin{equation}
\label{eq57}
\frac{\sigma^{2}}{\theta^{2}} = \frac{\{n(1 - \alpha) + (1 + \alpha)\}^
{2} - 3n(1 - \alpha^{2}) - 3\alpha }
{3\{n(1 - \alpha) + (1 + \alpha)\}^{2}} =  \mbox{constant}.
\end{equation}
The rotation $\omega$ is identically zero. \\

The rate of expansion $H_{i}$ in the direction of x, y and z are given by
$$
H_{x} = \frac{A_{4}}{A} = \frac{nb(1 - \alpha)}{(\ell + 1)}\frac{\psi_{1}}{\psi_{2}},
$$
$$
H_{y} = \frac{B_{4}}{B} = \frac{b}{(\ell + 1)}\frac{\psi_{1}}{\psi_{2}},
$$
\begin{equation}
\label{eq58}
H_{z} = \frac{C_{4}}{C} = \frac{b\alpha}{(\ell + 1)}\frac{\psi_{1}}{\psi_{2}}.
\end{equation}
Generally the model (\ref{eq31}) represents an expanding, shearing
and non-rotating universe in which the flow vector is geodetic. The
model (\ref{eq31}) starts expanding at $T > 0$ and goes on expanding
indefinitely when $\frac{n(1 - \alpha)}{(\ell + 1)} < 0$. Since
$\frac{\sigma}{\theta}$ = constant, the model does not approach
isotropy. As $T$ increases the proper volume also increases. The
physical quantities $p$ and $\rho$ decrease as $F_{12}$ increases.
However, if $\frac{n(1 - \alpha)}{(\beta + 1)} > 0$, the process of
contraction starts at $T>0$ and at $T = \infty$ the expansion stops.
The electromagnetic field tensor does not vanish when $b \ne 0$, and
$\alpha \ne 1$. It is observed from Eq. (\ref{eq55}) that $q < 0$
when $c_{1} > 0$ and $c_{2} > 0$ which implies an accelerating model
of the universe. Recent observations of type Ia supernovae \cite
{ref76,ref77} reveal that the present universe is in accelerating
phase and deceleration parameter lies somewhere in the range $-1 < q
\leq 0$. It follows that our models of the universe are consistent
with recent observations. Either when $c_{1} = 0$ or $c_{2} = 0$,
the deceleration parameter $q$ approaches the value $(-1)$ as in the
case of de-Sitter universe.
%%%%%%%%%%%%%%%%%%%%%%%%%%%%%%%%%%%%%%%%%%%%%%%%%%%%%%%%%%%%%%%%%%%%%%%%%
%%%%%%%%%%%%%%%%%%%%%%%%%%%%%%%  SECTION 6  %%%%%%%%%%%%%%%%%%%%%%%%%%%%%%%
\section{Solution in Absence of Magnetic Field}
In absence of magnetic field, the field Eq. (\ref{eq9}) with Eqs. (\ref{eq2}) and
(\ref{eq10}) for
metric (\ref{eq1}) read as
\begin{equation}
\label{eq59}
\frac{1}{A^{2}}\left[- \frac{B_{44}}{B} - \frac{C_{44}}{C} + \frac{A_{4}}{A}
\left(\frac{B_{4}}{B} +
\frac{C_{4}}{C}\right) - \frac{B_{4} C_{4}}{B C} + \frac{B_{1}C_{1}}{BC}\right]
= 8 \pi p + \frac{3}{4}\beta^{2} ,
\end{equation}
\begin{equation}
\label{eq60}
\frac{1}{A^{2}}\left(\frac{A^{2}_{4}}{A^{2}}- \frac{A_{44}}{A} - \frac{C_{44}}{C} +
\frac{C_{11}}{C} \right) =  8 \pi p + \frac{3}{4}\beta^{2},
\end{equation}
\begin{equation}
\label{eq61}
\frac{1}{A^{2}}\left(\frac{A^{2}_{4}}{A^{2}} - \frac{A_{44}}{A} - \frac{B_{44}}{B} +
\frac{B_{11}}{ B}\right) =  8 \pi p + \frac{3}{4}\beta^{2},
\end{equation}
\begin{equation}
\label{eq62}
\frac{1}{A^{2}}\left[- \frac{B_{11}}{B} - \frac{C_{11}}{C} + \frac{A_{4}}{A}
\left(\frac{B_{4}}{B} +
\frac{C_{4}}{C}\right) - \frac{B_{1}C_{1}}{BC}  + \frac{B_{4} C_{4}}{B C}\right]
= 8 \pi \rho - \frac{3}{4}\beta^{2},
\end{equation}
\begin{equation}
\label{eq63}
\frac{B_{14}}{B} + \frac{C_{14}}{C} - \frac{A_{4}}{A}\left(\frac{B_{1}}{B} +
\frac{C_{1}}{C}\right) = 0,
\end{equation}
Eqs. (\ref{eq60}) and (\ref{eq61}) lead to
\begin{equation}
\label{eq64}
\frac{B_{44}}{B} - \frac{B_{11}}{B} - \frac{C_{44}}{C} + \frac{C_{11}}{C} = 0.
\end{equation}
Eqs. (\ref{eq19}) and (\ref{eq64}) lead to
\begin{equation}
\label{eq65}
\frac{g_{44}}{g} - \frac{k_{44}}{k} = 0.
\end{equation}
Eqs. (\ref{eq23}) and (\ref{eq65}) lead to
\begin{equation}
\label{eq66}
\frac{g_{44}}{g} + \alpha \frac{g^{2}_{4}}{g^{2}} = 0,
\end{equation}
which on integration gives
\begin{equation}
\label{eq67}
g = (c_{4} t + c_{5})^{\frac{1}{(\alpha + 1)}},
\end{equation}
where $c_{4}$ and $c_{5}$ are constants of integration. Hence from (\ref{eq23}) and
(\ref{eq67}), we have
\begin{equation}
\label{eq68}
k = c(c_{4} t + c_{5})^{\frac{\alpha}{(\alpha + 1)}}.
\end{equation}
In this case (\ref{eq17}) leads to
\begin{equation}
\label{eq69}
f = \exp{\left(\frac{1}{2}K(x + x_{0})^{2}\right)}.
\end{equation}
Therefore, we have
\begin{equation}
\label{eq70}
B = \exp{\left(\frac{1}{2}K(x + x_{0})^{2}\right)} (c_{4} t + c_{5})^{\frac{1}
{(\alpha + 1)}},
\end{equation}
\begin{equation}
\label{eq71}
C = \exp{\left(\frac{1}{2}K(x + x_{0})^{2}\right)} c (c_{4} t + c_{5})^{\frac{\alpha}
{(\alpha + 1)}},
\end{equation}
\begin{equation}
\label{eq72}
A = a (c_{4} t + c_{5})^{\frac{n(1 - \alpha)}{(1 + \alpha)}},
\end{equation}
where $a$ is already defined in previous section. \\
After using suitable transformation of the co-ordinates, the metric (\ref{eq1})
reduces to the form
\[
ds^{2} = a^{2}(c_{4}T)^{\frac{2n(1 - \alpha)}{(1 + \alpha)}}(dX^{2} - dT^{2})
+ e^{KX^{2}}(c_{4} T)^{\frac{2}{(\alpha + 1)}} dY^{2}
\]
\begin{equation}
\label{eq73}
+ e^{KX^{2}}(c_{4} T)^{\frac{2\alpha}{(\alpha + 1)}} dZ^{2},
\end{equation}
where $x + x_{0} = X$, $y = Y$, $cz = Z$, $t + \frac{c_{5}}{c_{4}} = T$. \\

For the specification of displacement field $\beta(t)$ within the
framework of Lyra geometry and for realistic models of physical
importance, we consider the following two cases given in Sections $7$ and $8$.
%%%%%%%%%%%%%%%%%%%%%%%%%%%%%%%%%%%%%%%%%%%%%%%%%%%%%%%%%%%%%%%%%%%%%%%%%%%%%%%%%%%
%%%%%%%%%%%%%%%%%%%%%%%%%%%%%%%  SECTION 7  %%%%%%%%%%%%%%%%%%%%%%%%%%%%%%%%%%%%%%%%
\section{When $\beta$ is a constant i.e. $\beta = \beta_{0}$ (constant)}
Using Eqs. (\ref{eq70})-(\ref{eq72}) in Eqs.\ref{eq59}) and (\ref{eq62}) the 
expressions for pressure $p$ and density $\rho$ for the model (\ref{eq73}) 
are given by
\begin{equation}
\label{eq74}
8 \pi p = \frac{1}{a^{2}(c_{4}T)^{\frac{2n(1 - \alpha)}{(1 + \alpha)}}}
\Biggl[\left\{\frac{n(1 - \alpha^{2}) + \alpha}{(\alpha + 1)^{2}}\right \}
\frac{1}{T^{2}} + K^{2}X^{2}\Biggr] - \frac{3}{4}\beta_{0}^{2},
\end{equation}
\begin{equation}
\label{eq75}
8 \pi \rho = \frac{1}{a^{2}(c_{4}T)^{\frac{2n(1 - \alpha)}{(1 + \alpha)}}}
\Biggl[\left\{\frac{n(1 - \alpha^{2}) + \alpha}{(\alpha + 1)^{2}}\right \}
\frac{1}{T^{2}} - K(2 + 3KX^{2})\Biggr] +  \frac{3}{4}\beta_{0}^{2},
\end{equation}
The dominant energy conditions (Hawking and Ellis \cite{ref75})
$$
(i) \rho - p \geq 0, ~ ~ (ii) \rho + p \geq 0,
$$
lead to
\begin{equation}
\label{eq76}
\frac{3}{4}\beta_{0}^{2}a^{2}(c_{4}T)^{\frac{2n(1 - \alpha)}{(1 + \alpha)}}
\geq K(1 + 2KX^{2}),
\end{equation}
and
\begin{equation}
\label{eq77}
\left\{\frac{n(1 - \alpha^{2}) + \alpha}{(1 + \alpha)^{2}}\right\}
\frac{1}{T^{2}} \geq K(1 + KX^{2}).
\end{equation}
respectively.

The reality conditions (Ellis \cite{ref74})
$$
(i) \rho + p > 0, ~ ~ (ii) \rho + 3p > 0,
$$
lead to
\begin{equation}
\label{eq78}
\left\{\frac{n(1 - \alpha^{2}) + \alpha}{(1 + \alpha)^{2}}\right\}\frac{1}{T^{2}} > K(1 + KX^{2}),
\end{equation}
and
\begin{equation}
\label{eq79}
\frac{2[n(1 - \alpha^{2}) + \alpha]}{(1 + \alpha)^{2}}\frac{1}{T^{2}} > K +
\frac{3}{4}\beta_{0}^{2}
(c_{4}T)^\frac{2n(1 - \alpha)}{(1 + \alpha)}.
\end{equation}
The condition (\ref{eq76}) and (\ref{eq79}) impose a restriction on $\beta_{0}$.
%%%%%%%%%%%%%%%%%%%%%%%%%%%%%%%%%%%%%%%%%%%%%%%%%%%%%%%%%%%%%%%%%%%%%%%%%%%%%%%%%%%
%%%%%%%%%%%%%%%%%%%%%%%%%%%%%%%  SECTION 8  %%%%%%%%%%%%%%%%%%%%%%%%%%%%%%%%%%%%%%%%
\section{When $\beta$ is a function of $t$}
In this case to find the explicit value of displacement field
$\beta(t)$, we assume that the fluid obeys an equation of state
given by (\ref{eq39}). Using Eqs. (\ref{eq70}) - (\ref{eq72}) 
and (\ref{eq39}) in Equations (\ref{eq59}) and (\ref{eq62}) we obtain 
expressions for $\rho(t)$ and $\beta{(t)}$ given by
\begin{equation}
\label{eq80}
8 \pi (1 + \gamma)\rho = \frac{1}{a^{2}(c_{4}T)^{\frac{2n(1 - \alpha)}{(1 + \alpha)}}}
\Biggl[\left\{\frac{n(1 - \alpha^{2}) + \alpha}{(\alpha + 1)^{2}}\right \}
\frac{2}{T^{2}} - 2K(1 + KX^{2})\Biggr],
\end{equation}
\[
(1 + \gamma)\beta^{2}{(t)}  = \frac{4}{3a^{2}(c_{4}T)^{\frac{2n(1 - \alpha)}{(1 + \alpha)}}}
\Biggl[\left\{\frac{n(1 - \alpha^{2}) + \alpha}{(\alpha + 1)^{2}}\right \}
\frac{(1 - \gamma)}{T^{2}} 
\]
\begin{equation}
\label{eq81}
+ 2K \gamma + KX^{2} (1 + 3\gamma)\Biggr].
\end{equation}
It is observed that $\rho > 0$ and $\beta^{2}{(t)} > 0$ according as
\begin{equation}
\label{eq82}
\left\{\frac{n(1 - \alpha^{2}) + \alpha}{(\alpha + 1)^{2}}\right \}
\frac{1}{T^{2}} > K(1 + KX^{2}),
\end{equation}
and
\begin{equation}
\label{eq83}
\left\{\frac{n(1 - \alpha^{2}) + \alpha}{(\alpha + 1)^{2}}\right \}
\frac{1}{T^{2}} > K^{2} X^{2},
\end{equation}
respectively.
 
It is worth mention here that by putting $\gamma = 0, 1, \frac{1}{3}$ in 
Eqs. (\ref{eq80}) and (\ref{eq81}), one can derive the expressions for energy 
density $\rho(t)$ and  displacement vector $\beta{(t)}$ for empty universe, 
Zeldovich universe and radiating universe respectively. It is also observed 
that these three types of models have similar properties as we have already 
discussed above. Therefore, we have not mentioned the expressions for physical 
quantities of these models.\\

%%%%%%%%%%%%%%%%%%%%%%%%%%%%%%%%%%%%%%%%%%%%%%%%%%%%%%%%%%%%%%%%%%%%%%%%

\noindent
{\bf{Some Geometric Properties of the Model}} \\
The expressions for the expansion $\theta$, Hubble parameter $H$,
shear scalar $\sigma^{2}$, deceleration parameter $ q $ and proper
volume $V^{3}$ for the model (\ref{eq73}) in absence of magnetic
field are given by
\begin{equation}
\label{eq84}
\theta = 3H = \frac{n(1 - \alpha) + (1 + \alpha)}{a(1 +
\alpha)c_{4}^{\frac{n(1 - \alpha)} {(1 + \alpha)}}}
\frac{1}{T^{\frac{n(1 - \alpha) + (1 + \alpha)}{(1 + \alpha)}}}
\end{equation}
\begin{equation}
\label{eq85}
\sigma^{2} =  \frac{\{n(1 - \alpha) + (1 + \alpha)\}^{2} - 3n(1 - \alpha^{2}) - 3\alpha}
{3 a^{2}(1 + \alpha)^{2}c_{4}^{\frac{n(1 - \alpha)}{(1 + \alpha)}}}
\frac{1}{T^{\frac{2n(1 - \alpha) + 2(1 + \alpha)}{(1 + \alpha)}}}
\end{equation}
\begin{equation}
\label{eq86}
q = - 1 + \frac{3(\alpha + 1)}{2n(1 - \alpha) + 2(1 + \alpha)},
\end{equation}
\begin{equation}
\label{eq87}
V^{3} = \sqrt{-g} = a^{2} e^{KX^{2}} (c_{4}T)^{\frac{2n(1 - \alpha) + (1 + \alpha)}
{(1 + \alpha)}}.
\end{equation}
From Eqs. (\ref{eq84}) and (\ref{eq85}) we obtain
\begin{equation}
\label{eq88}
\frac{\sigma^{2}}{\theta^{2}} = \frac{\{n(1 - \alpha) + (1 + \alpha)\}^{2} -
3n(1 - \alpha^{2}) - 3\alpha }
{3\{n(1 - \alpha) + (1 + \alpha)\}^{2}} =  \mbox{constant}.
\end{equation}
The rotation $\omega$ is identically zero. \\
The rate of expansion $H_{i}$ in the direction of x, y and z are given by
$$
H_{x} = \frac{A_{4}}{A} = \frac{n(1 - \alpha)}{(1 + \alpha)}\frac{1}{T},
$$
$$
H_{y} = \frac{B_{4}}{B} = \frac{1}{(1 + \alpha)}\frac{1}{T},
$$
\begin{equation}
\label{eq89}
H_{z} = \frac{C_{4}}{C} = \frac{\alpha}{(1 + \alpha)}\frac{1}{T}.
\end{equation}
The model (\ref{eq73}) starts expanding with a big bang at $T = 0$
and it stops expanding at $T = \infty$. It should be noted that the
universe exhibits initial singularity of the Point-type at $T = 0$.
The space-time is well behaved in the range $0 < T < T_{0}$. In
absence of magnetic field the model represents a shearing and
non-rotating universe in which the flow vector is geodetic. At the
initial moment $T = 0$, the parameters $\rho$, $p$, $\beta$,
$\theta$, $\sigma^{2}$ and $H$ tend to infinity. So the universe
starts from initial singularity with infinite energy density,
infinite internal pressure, infinitely large gauge function,
infinite rate of shear and expansion. Moreover, $\rho$, $p$,
$\beta$, $\theta$, $\sigma^{2}$ and $H$ are monotonically decreasing
toward a non-zero finite quantity for $T$ in the range $0 < T <
T_{0}$ in absence of magnetic field. Since $\frac{\sigma}{\theta}$ =
constant, the model does not approach isotropy. As $T$ increases the
proper volume also increases. It is observed that for all the three
models i.e. for empty universe, Zeldovice universe and radiating
universe, the displacement vector $\beta(t)$ is a decreasing
function of time and therefore it behaves like cosmological term
$\Lambda$. It is observed from Eq. (\ref{eq86}) that $q < 0$ when
$\alpha < \frac{2n - 1}{2n + 1}$ which implies an accelerating model
of the universe. When $ \alpha = -1$, the deceleration parameter $q$
approaches the value $(-1)$ as in the case of de-Sitter universe.
Thus, also in absence of magnetic field, our models of the universe
are consistent with recent observations.
%%%%%%%%%%%%%%%%%%%%%%%%%%%%%%%%%%%%%%%%%%%%%%%%%%%%%%%%%%%%%%%%%%%%%%%%%%%%%%%%%%%
%%%%%%%%%%%%%%%%%%%%%%%%%%%%%%%  SECTION 9  %%%%%%%%%%%%%%%%%%%%%%%%%%%%%%%%%%%%%%%%
\section{Discussion and Concluding Remarks}
In this paper, we have obtained a new class of exact solutions of
Einstein's field equations for cylindrically symmetric space-time
with perfect fluid distribution within the framework of Lyra's
geometry both in presence and absence of magnetic field. The
solutions are obtained using the functional separability of the
metric coefficients. The source of the magnetic field is due to an
electric current produced along the z-axis. $F_{12}$ is the
only non-vanishing component of electromagnetic field tensor. The
electromagnetic field tensor is given by equation (\ref{eq34}),
$\bar{\mu}$ remains undetermined as function of both $x$ and $t$.
The electromagnetic field tensor does not vanish if $b \ne 0$ and
$\alpha \ne 1$. It is observed that in presence of magnetic field,
the rate of expansion of the universe is faster than that in absence
of magnetic field. The idea of primordial magnetism is appealing
because it can potentially explain all the large-scale fields seen
in the universe today, specially those found in remote
proto-galaxies. As a result, the literature contains many studies
examining the role and the implications of magnetic fields for
cosmology. In presence of magnetic field the  model (\ref{eq31})
represents an expanding, shearing and non-rotating universe in which
the flow vector is geodetic. But in the absence of magnetic field
the model (\ref{eq70}) is found that in the universe all the matter
and radiation are concentrated at the big bang epoch and the cosmic
expansion is driven by the big bang impulse. The universe has
singular origin and it exhibits power-law expansion after the big
bang impulse. The rate of expansion slows down and finally stops at
$T \to \infty$. In absence of magnetic field, the pressure, energy
density and displacement field become zero whereas the spatial
volume becomes infinitely large as $T \to \infty$.
\par
It is possible to discuss entropy in our universe. In thermodynamics
the expression for entropy is given by
\begin{equation}
\label{eq90} TdS = d(\rho V^{3}) + p(dV^{3}),
\end{equation}
where $V^{3} = A^{2}BC$ is the proper volume in our case. To solve
the entropy problem of the standard model, it is necessary to treat
$dS > 0$ for at least a part of evolution of the universe. Hence Eq.
(\ref{eq90}) reduces to
\begin{equation}
\label{eq91} TdS = \rho_{4} + (\rho + p)\left(2\frac{A_{4}}{A} +
\frac{B_{4}}{B} + \frac{C_{4}} {C}\right) > 0.
\end{equation}
The conservation equation $T^{j}_{i:j} = 0$ for (\ref{eq1}) leads to
\begin{equation}
\label{eq92} \rho_{4} + (\rho + p)\left(\frac{A_{4}}{A} +
\frac{B_{4}}{B} + \frac{C_{4}} {C}\right) + \frac{3}{2}\beta
\beta_{4} + \frac{3}{2}\beta^{2}\left(2\frac{A_{4}}{A} +
\frac{B_{4}}{B} + \frac{C_{4}}{C}\right) = 0.
\end{equation}
Therefore, Eqs. (\ref{eq91}) and (\ref{eq92}) lead to
\begin{equation}
\label{eq93} \frac{3}{2}\beta \beta_{4} +
\frac{3}{2}\beta^{2}\left(2\frac{A_{4}}{A} + \frac{B_{4}}{B} +
\frac{C_{4}}{C}\right) < 0.
\end{equation}
which gives to $\beta < 0$. Thus, the displacement vector $\beta(t)$
affects entropy because for entropy $dS > 0$ leads to $\beta(t) <
0$.
\par
In spite of homogeneity at large scale our universe is inhomogeneous
at small scale, so physical quantities being position-dependent are
more natural in our observable universe if we do not go to super
high scale. This result shows this kind of physical importance. It
is observed that the displacement vector $\beta(t)$ coincides with
the nature of the cosmological constant $\Lambda$  which has been
supported by the work of several authors as discussed in the
physical behaviour of the model in Sections $5$ and $8$. In recent
time $\Lambda$-term has attracted theoreticians and observers for
many a reason. The nontrivial role of the vacuum in the early
universe generates a $\Lambda$-term that leads to inflationary
phase. Observationally, this term provides an additional parameter
to accommodate conflicting data on the values of the Hubble
constant, the deceleration parameter, the density parameter and the
age of the universe (for example, see Refs. \cite{ref78} and
\cite{ref79}). Assuming that $\Lambda$ owes its origin to vacuum
interaction, as suggested particularly by Sakharov \cite{ref80}, it
follows that it would, in general, be a function of space and time
coordinates, rather than a strict constant. In a homogeneous
universe $\Lambda$ will be at most time dependent \cite{ref81}. In
the case of inhomogeneous universe this approach can generate
$\Lambda$ that varies both with space and time. In considering the
nature of local massive objects, however, the space dependence of
$\Lambda$ cannot be ignored. For details, reference may be made to
Refs. \cite{ref82}, \cite{ref83}, \cite{ref84}. In recent past there
is an upsurge of interest in scalar fields in general relativity and
alternative theories of gravitation in the context of inflationary
cosmology \cite{ref85,ref86,ref87}. Therefore the study of
cosmological models in Lyra's geometry may be relevant for
inflationary models. Also the space dependence of the displacement
field $\beta$ is important for inhomogeneous models for the early
stages of the evolution of the universe. In the present study we
also find $\beta(t)$ as both space and time dependent which may be
useful for a better understanding of the evolution of universe in
cylindrically symmetric space-time within the framework of Lyra's
geometry. There seems a good possibility of Lyra's geometry to
provide a theoretical foundation for relativistic gravitation,
astrophysics and cosmology. However, the importance of Lyra's
geometry for astrophysical bodies is still an open question. In
fact, it needs a fair trial for experiment.
\section*{Acknowledgements}
\noindent The authors would like to thank the Harish-Chandra
Research Institute, Allahabad, India for local hospitality where
this work is done.
\newline
%% \newpage
%\nonumsection{References}

\end{document}